\documentclass[twocolumn,superscriptaddress,showpacs,amsmath,amssymb,pra,longbibliography]{revtex4-1}

\usepackage[pdftex]{graphicx} 
\usepackage{dcolumn}  
\usepackage{bm}       
\usepackage[usenames,dvipsnames]{color}
\definecolor{URLCOL}{rgb}{0,0.52,0.83} 
\definecolor{LINKCOL}{rgb}{0.05,0.5,0} 
\definecolor{orange}{rgb}{0.6,0.3,0} 
\definecolor{CITECOL}{rgb}{0.25,0,0.48} 
\usepackage{epstopdf}

\usepackage[pdftex,bookmarks,breaklinks,bookmarksopen,bookmarksnumbered,colorlinks,linkcolor=LINKCOL,linktocpage,citecolor=CITECOL,urlcolor=URLCOL,pdfpagemode=UseOutline,pdftex]{hyperref}

\usepackage{fancyhdr}
\fancypagestyle{titlepage}
{\chead{file: \jobname}\rhead{Total pages: \pageref{page:end}}}

\newcommand{\bd}[1]{\boldsymbol{#1}}

\usepackage{booktabs}
\usepackage{natbib}

\definecolor{TITLECOL}{rgb}{0.1,0.2,0.7} 
\definecolor{SECOL}{rgb}{0.1,0.2,0.7} 
\definecolor{CONTENTSCOL}{rgb}{0.1,0.2,0.7} 
\definecolor{SSECOL}{rgb}{0.25,0,0.48} 
\definecolor{SSSECOL}{rgb}{0.2,0.08,0.53} 
\definecolor{FINCOL}{rgb}{0.01,0.3,0.07} 

\def\coloredtitle#1{\title{\textcolor{TITLECOL}{#1}}} 
\def\coloredauthor#1{\author{\textcolor{CITECOL}{#1}}} 

\definecolor{URLCOL}{rgb}{0,0.17,0.43} 
\definecolor{LINKCOL}{rgb}{0.05,0.4,0} 
\definecolor{CITECOL}{rgb}{0.35,0,0.48} 

\def\sss{\scriptscriptstyle\rm}

\def\bea{\begin{eqnarray}}
\def\eea{\end{eqnarray}}
\def\ben{\begin{equation}}
\def\een{\end{equation}}
\def\benu{\begin{enumerate}}
\def\enu{\end{enumerate}}

\def\bei{\begin{itemize}}
\def\eei{\end{itemize}}
\def\beit{\begin{itemize}}
\def\eit{\end{itemize}}
\def\benu{\begin{enumerate}}
\def\enu{\end{enumerate}}

\def\br{{\bf r}}

\def\x{_{\sss X}}
\def\c{_{\sss C}}
\def\s{_{\sss S}}

\def\H{_{\sss H}}

\def\n{n}

\begin{document}

\coloredtitle{
Quantum critical benchmark for density functional theory
}
\coloredauthor{Paul E. Grabowski}
\email{paul.grabowski@uci.edu}
\affiliation{Department of Chemistry, University of California, Irvine, CA 92697, USA}

\coloredauthor{Kieron Burke}

\email{kieron@uci.edu}
\affiliation{Departments of Chemistry and Physics, University of California, Irvine, CA 92697, USA}

\begin{abstract}
Two electrons at the threshold of ionization represent a severe test case
for electronic structure theory. 
A pseudospectral method yields a very accurate density of
the two-electron ion with nuclear charge close to the critical value.
Highly accurate energy components and potentials of Kohn-Sham
density functional theory are given, as well
as a useful parametrization of the critical density.
The challenges for density functional approximations and the
strength of correlation are also discussed.
\end{abstract}

\maketitle

The value of highly accurate benchmark calculations to first-principles electronic structure theory
cannot be overstated.   While comparison with experiment is the ultimate arbiter of the usefulness
of prediction, the ability to control and eliminate multiple sources of error with a direct solution of 
the Schr\"{o}dinger equation allows pure
`apples-to-apples' comparisons.  These have proven invaluable in the development of
Kohn-Sham (KS) density functional theory (DFT)\,\cite{KS65}, where the methodology is so alien to standard
wave function treatments that usually only a detailed comparison of ground-state energies can be
used to test approximations.   
For a recent example, van der Waals theories
use highly accurate quantum chemical methods on small molecules to 
validate (or not) approximate functionals, not just at equilibrium bond lengths, but
for entire binding energy curves\,\cite{KM12}. 
Accurate energies and densities can
also be used to distinguish energy-driven from density-driven
errors in DFT\,\cite{KSB13}.

Beyond ground-state energy comparisons, 
various energy components and potentials can 
be examined, once a sufficiently accurate density is available
from the benchmark calculation.  The pioneering work of Umrigar and coworkers\,\cite{UG94,HU97} 
for several spherical
atoms is a case in point.  The availability of the KS potential and its eigenvalues
was useful for all DFT, and especially for the development of linear-response time-dependent
density functional methods for finding excited state energies, where the ground-state orbitals
and energies are vital inputs\,\cite{AGB03}. 
A similar role was played by Baerends and coworkers for the
H$_2$ molecule as a function of bond length\,\cite{BBS89}.  This system has since become the paradigm of
strong correlation\,\cite{MCY08}.

The benchmark we consider here is, in some ways, the most fundamental to electronic structure
theory: two electrons bound to a single nucleus.   But we study the very special case when
the ionization potential is precisely zero, i.e., the nuclear charge $Z_c$ is the smallest
possible value that binds two electrons.  Thus the density is the most diffuse of any single-center
electronic system, making it extremely difficult to extract from most
methods.  For example, traditional quantum chemical basis sets fall off too rapidly at large $r$ to
extract the density, even if the energy is extremely accurate.  
This is the simplest case of a quantum-critical electronic problem\,\cite{SK96}.
Such systems have been mapped to 
phase transitions in statistical mechanics\,\cite{NSK97}.

Recently, high-precision variational calculations have
greatly expanded the accuracy to which $Z_c$ is known\,\cite{EBMD14}, and
esoteric strong correlation methods have been tested on this system\cite{MUMG14}.
Our work shows how pseudospectral methods are ideally suited for extracting
expectation values for 
weakly bound systems, demonstrated by an extremely accurate density at $Z_c$.
We parametrize this density in a simple form,
give the asymptotic density at large $r$ to order $r^{-4}$ and
highly accurate KS energy components,
and show the performance of popular DFT approximations.

\begin{figure}[!tbp]
\includegraphics[width=1\linewidth]{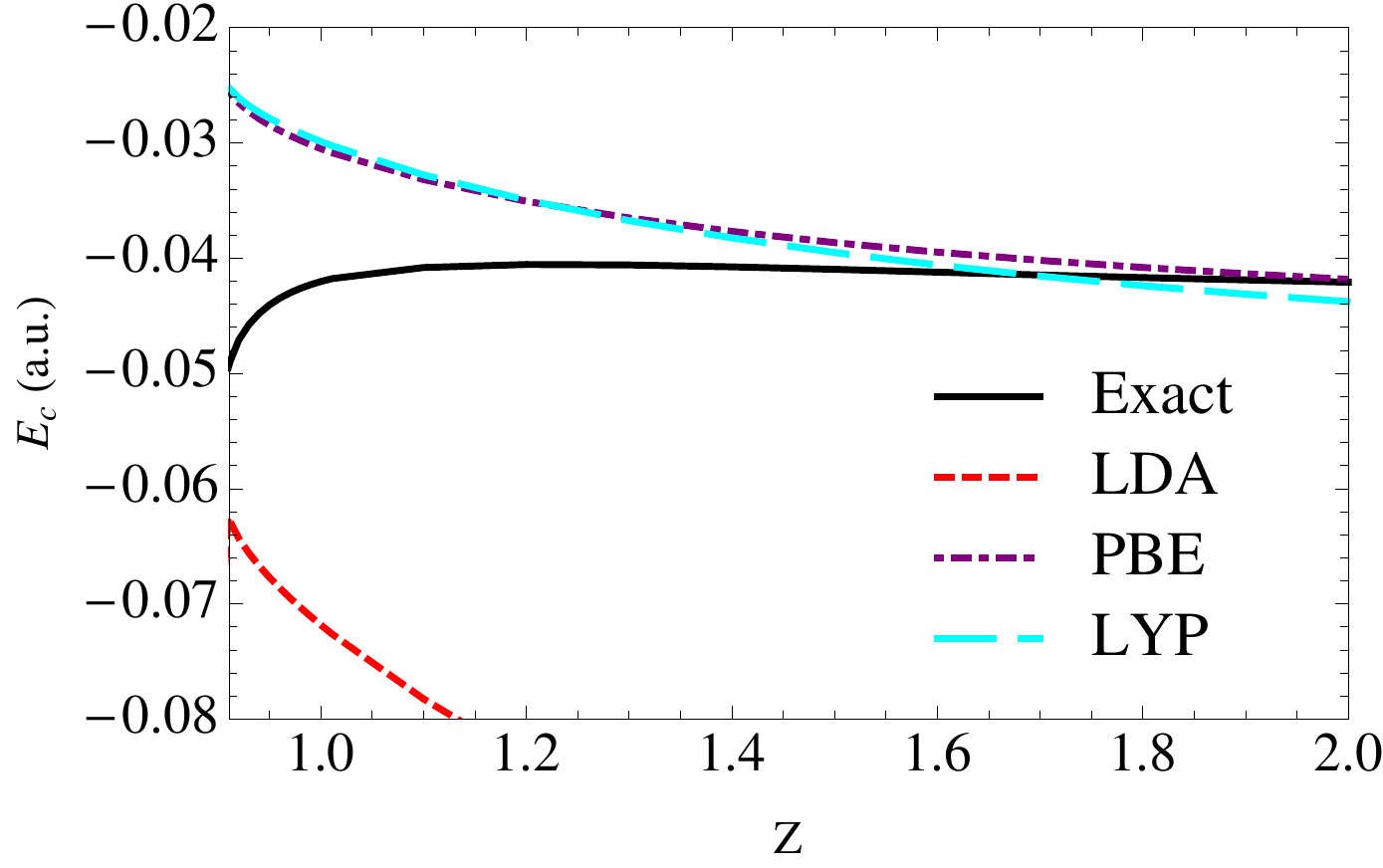}
\caption{\label{Ec} Exact and approximate correlation energies,
evaluated with the exact densities.  
}
\end{figure}
The usefulness of our benchmark is illustrated in Fig. \ref{Ec}.
Previous calculations\,\cite{UG94} for two-electron ions run from $Z=80$ (Hg$^{78+}$)
down to $Z=1$ (H$^-$).  The correlation energy is almost independent
of $Z$, and is roughly accounted for by modern approximations down
to $Z=2$ (He).   But the slope of $E\c(Z)$ changes sign below 2, an effect completely missed by
the commonly used functionals, PBE\,\cite{PBE96} and LYP\,\cite{LYP88,MSSP89}, 
which behave more poorly as the density becomes
more diffuse.   
(But see the end of this letter, where
this apparent catastrophe for modern approximations
mysteriously becomes a triumph!).

For many $N$-electron atoms, there exists a minimum $Z_c=N-1-\nu$, with $0<\nu<1$, 
such that the ground state of $\hat{H}(Z)$ has positive ionization energy for all nuclear charges $Z>Z_c$. 
It is thought that $\lambda_c=1/Z_c$ corresponds to the radius of convergence of 
the perturbative solution of the two-electron atom with the perturbation being the electron-electron interaction 
$1/r_{12}$\,\cite{BakerEtAl1990}.
Baker and coworkers used $401$ orders of perturbation theory to obtain $Z_c = 0.911\,03$\,\cite{BakerEtAl1990} and
Ivanov later used better extrapolation techniques on their data to get $Z_c=0.911\,028\,26$\,\cite{Ivanov1995}.
From a direct variational calculation to solve for the critical charge, Sergeev and Kais obtained $Z_c =0.911\,028\,225$\,\cite{SK99}.
Recently, Estienne and coworkers\,\cite{EBMD14} obtained $Z_c=0.911\,028\,224\,077\,255\,73(4)$, far surpassing
the prior estimates in precision. In the present work, we obtain $Z_c=0.911\,028\,224\,07(6)$, agreeing with Ref.\,\cite{EBMD14}
and an unpublished figure by Schwartz\,\cite{S13}. Although our critical
charge is not as precise, our wave function is roughly as accurate as 
our value of $Z_c$, allowing
the calculation of much more precise expectation values than a variational calculation\,\cite{GrabowskiChernoff2011}.

Standard quantum methods typically have much trouble calculating states near the ionization threshold.
Such difficulties stem from two reasons: an improper representation of the wave function near the 
electron-electron coalescence point and the mixing of energetically similar continuum states into the ground state
when using an approximate method.
For example, diffusion Monte Carlo calculations take advantage of the separation in energy between the 
ground state and excited states and fails to separate degenerate states. 
However, the pseudospectral method is a non-variational collocation method in which the value of the 
wave function is calculated on a grid in such a way that the {\it local} error in the wave function becomes exponentially small with increasing grid resolution.  
It allows us to accurately calculate the bound state right on the threshold of the continuum
by automatically selecting normalizable states. 

Pseudospectral methods\,\cite{NumericalRecipes} have their origins in fluid dynamics\,\cite{CanutoEtAl1988},
for which they are used to evolve systems without shocks
because their convergence properties only hold for $\mathcal{C}^\infty$ functions.
They have been extended to solving Einstein's field equations for colliding black holes
by the excision of the singularities from the computational domain
\cite{KidderFinn2000,PfeifferEtAl2003}. In quantum chemistry, Friesner
and others have 
shown orders of magnitude improvement in speed for a wide variety of methods
\cite{Friesner1985,Friesner1986,Friesner1987,RingnaldaEtAl1990,GreeleyEtAl1994,MurphyEtAl1995,MurphyEtAl2000,KoEtAl2008,HeylThirumalai2009}.
Direct solution of Schr\"{o}dinger's equation has been done for one-electron problems 
\cite{Borisov2001,BoydEtAl2003}, but only recently has
a sufficient representation of the computational domain been demonstrated for the case of fully-correlated, 
multi-electron atoms\,\cite{GrabowskiChernoff2010,GrabowskiChernoff2011}.  
Here we use the same implementation as in Ref.\,\cite{GrabowskiChernoff2011}.

\begin{figure}[!tbp]
\includegraphics[width=1\linewidth]{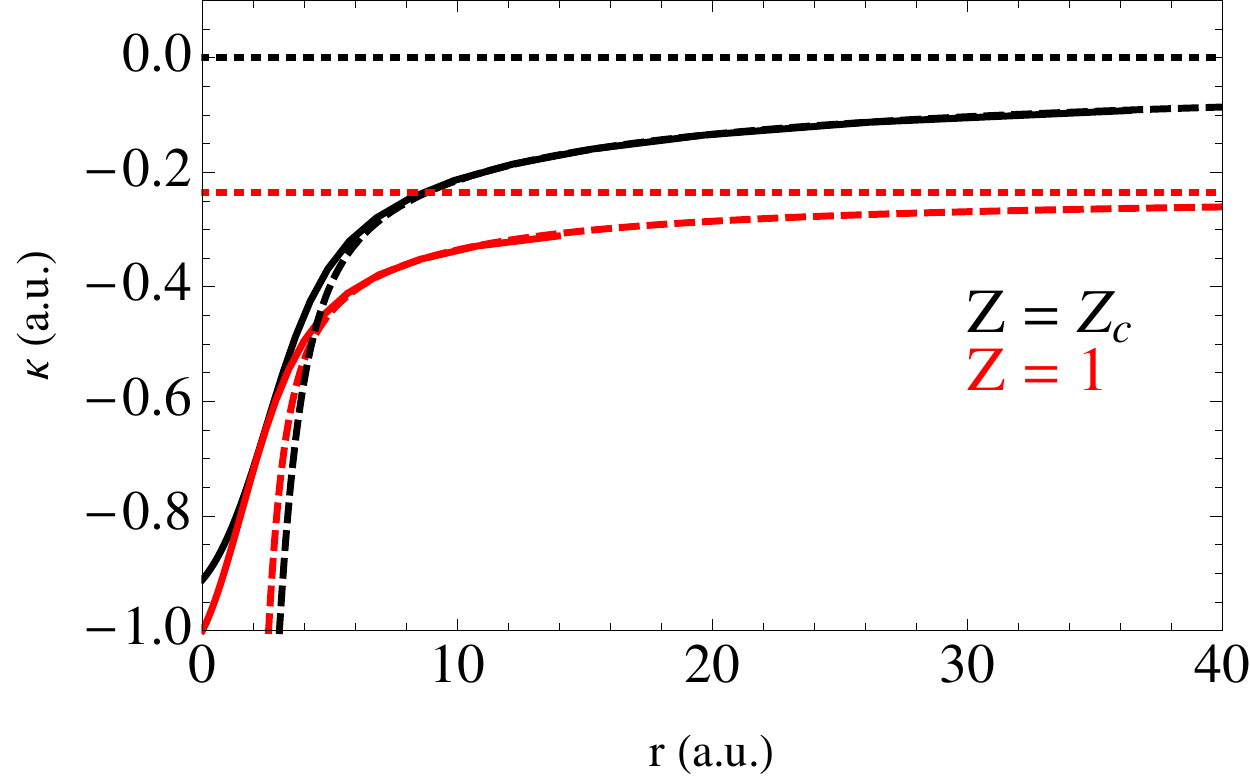}
\caption{\label{KappaPlot} Logarithmic derivative of the density ($2\kappa = d\ln n/dr$) as a function of 
$r$ for $Z = Z_c$ (upper, black) and $Z=1$ (lower, red). Solid lines are calculated.
The limits at large $r$ are shown as dotted lines and 
results of using Eqs. (\ref{NormalExpansion}) and (\ref{ZcExpansion}) are shown as the dashed lines.
}
\end{figure}

To illustrate the strength of the local convergence property, we plot in Fig. \ref{KappaPlot},
$\kappa=(1/2)d\log(\n)/dr$, where $n(r)$ is the one-electron density,
for $Z=Z_c$ and $Z=1$ (H$^-$).  
As $r\to \infty$, for $Z > Z_c$, the
well-known analysis of the exponential decay
of the density\,\cite{PPLB82} yields $\kappa \to -\sqrt{2I}$,
where $I$ is the ionization energy. However, for $Z=Z_c$, the behavior
differs qualitatively ($\kappa\to -2\sqrt{2(1-Z_c)/r}$). For both values of $Z$,
the asymptotic value is not approached until very large $r$.
Even at $r=40$ Bohr, there is a visible deviation in $\kappa(r)$ from its limits,
so one must use higher order expansions to connect the limits with our numeric results.

To analyze these results, we review well-known facts from KS DFT\cite{DG90}.
The KS equations describe fictitious non-interacting fermions sitting
in a potential, $v\s(\br)$, whose density matches the real one.  For two
spin-unpolarized electrons, one orbital is doubly occupied and the
KS equation in atomic units is
\ben
\label{kseq}
\left[-\frac{1}{2}\nabla^2+v_s(\mathbf{r})\right]\phi(\mathbf{r})=\epsilon\phi(\mathbf{r}),
\een
where $\phi=\sqrt{n(r)/2}$ and $\epsilon=-I=E+Z^2/2$ are the KS
orbital and its energy, respectively,
while $v_s(r)$ is the Kohn-Sham potential given by
\ben
v_s(\mathbf{r})=-\frac{Z}{r}+v\H(\mathbf{r})+v\x(\mathbf{r})+v\c(\mathbf{r}).
\een
For two electrons, the Hartree and exchange potentials are trivially related,
\ben
v\H(\mathbf{r})=-2\, v\x(\mathbf{r})=\int\frac{n(\mathbf{r}')d\mathbf{r}'}{|\mathbf{r}-\mathbf{r}'|}
\een
and the correlation potential is defined so as to make Eq. (\ref{kseq}) exact.
For large $r$, the exchange potential, behaves
as $-1/r$
while the correlation potential decays much faster, as $-\alpha/2r^4$, where
$\alpha$ is the dipole polarizability of the $N-1$ system, here equal to $9/2Z^4$\,\cite{UG94}.
Amovilli and March\,\cite{AmovilliMarch2006} derived the asymptotic behavior of the density
at large $r$ for $Z=2$. Here, we extend
their work to any $Z>Z_c$ and to the next highest order in $1/r$:
\ben
\label{NormalExpansion}
{\sqrt{n_Z(r)}}\sim \frac{x^\beta \sqrt{A}}{e^{x}}\left[\sum_{k=0}^4\frac{a_k}{x^k}-\frac{3r^{-2}}{4 Z^4}\sum_{k=1}^2\frac{\tilde{a}_k}{x^k}+
\mathcal{O}\left(x^{-5}\right)\right],
\een
where $\delta=Z-Z_c$, 
$x=\eta\, r$, $\eta=\sqrt{2 I}$, $\beta=\xi/\eta -1$, and $\xi=Z-1$.  The constant $A$ is defined as
$A=\lim_{r\to\infty} {n_Z(r)}\,e^{2x}/x^{2\beta}$.
The formula $A = \alpha_1\delta+\alpha_2 \delta^2+\alpha_3\delta^3$, with
$\alpha_1 = 0.006\,674\,48$, $\alpha_2=0.567\,102$, and $\alpha_3=2/\pi$ fits our 
densities over the entire $Z$ range to within $0.2\%$ with the maximum error occurring around $Z=1$.
The value of $\alpha_3$ comes from the large-$Z$ limit of the density.
Likewise, $I = \{1 + \beta_1 \exp[-\beta_2 \ln^2(\beta_3 \delta)]\}(\beta_4\delta+\delta^2/2)$,
with $\beta_1=0.085\,704\,8$, $\beta_2 = 0.166\,941$, $\beta_3 = 5.097\,253$, and $\beta_4 = 2\langle 1/r_1\rangle_{Z=Z_c}-Z_c=0.245\,189\,01$ has a maximum error of $0.3\%$ occurring around $Z=0.92$.
The coefficients in the large-$r$ expansion are given recursively as 
$a_0=1$, $a_k=-a_{k-1}[\xi-k\eta][\xi-(k-1)\eta]/(2k\eta^2)$, and
$\tilde{a}_0=1$ and
$\tilde{a}_1=-(\beta^2+\beta+3)/2$.

\begin{figure}[!tbp]
\includegraphics[width=1\linewidth]{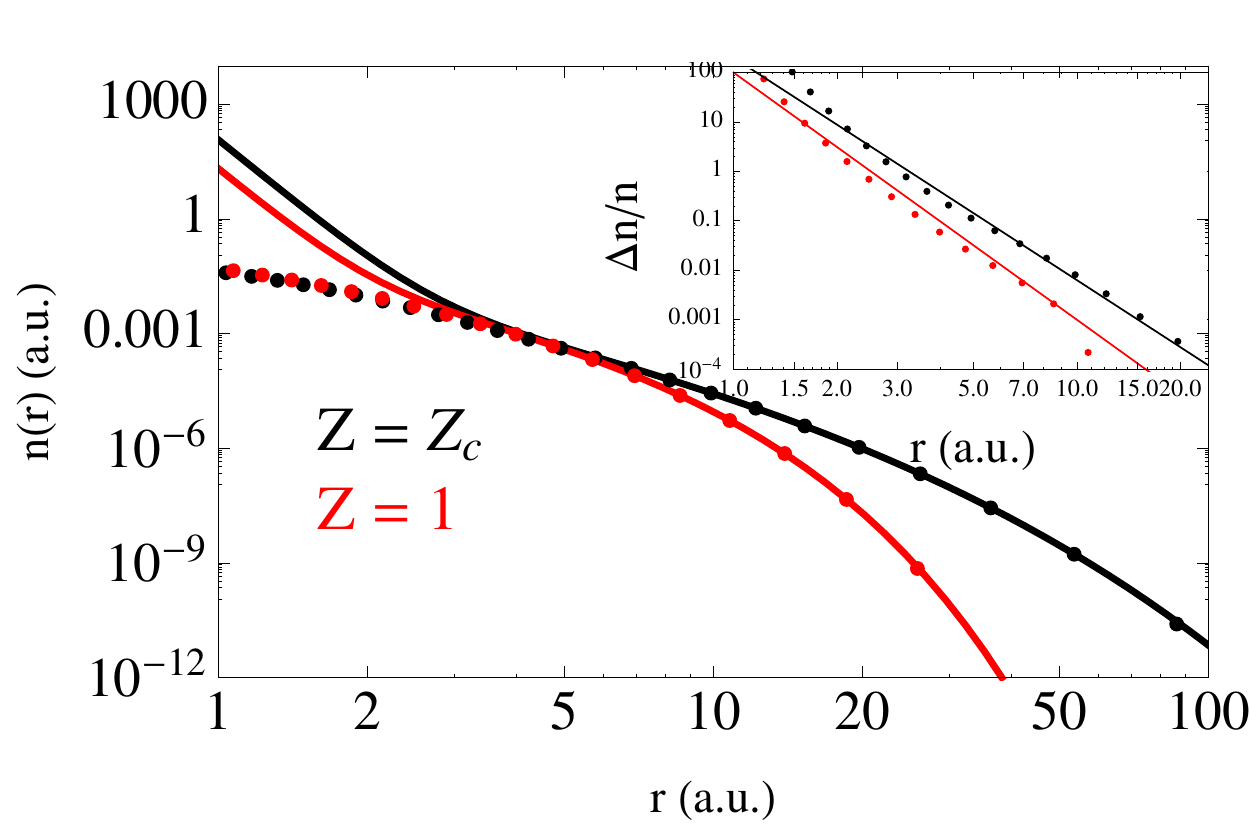}
\caption{\label{DensNAsym} Exact (dots) and asymptotic ($A=0.005528$ and $B=0.1375$) densities (solid line) for $Z=Z_c$ 
(upper, black) and $Z=1$ (lower, red). The fractional error (dots) in the asymptotic forms is shown in the inset along with  curves 
proportional to $r^{-4.5}$ (upper, black) and $r^{-5}$ (lower, red).} 
\end{figure}

At $Z_c$, the long range behavior
changes\,\cite{HHS83,AmovilliMarch2006,MUMG14}. 
Here we extend such asymptotic forms to higher order:
\ben
\label{ZcExpansion}
{\sqrt{n_{Z_c}(r)}} \sim  \frac{\sqrt{B} e^{-y}}{r^{3/4}}
\left[\sum_{k=0}^8\frac{b_k}{y^k}
- \frac{9r^{-2}}{4Z_c^4}\sum_{k=1}^4\frac{\tilde{b}_k}{y^{k}}+\mathcal{O}
\left({y^{-9}}\right)\right],
\een
where $B=\lim_{r\to\infty}n_{Z_c}(r)e^{2y} y^{3/2}\approx 0.1375$, $y=2\sqrt{2|\xi_c| r}$, $\xi_c=Z_c-1$, and
$b_0 =1$, $b_k=-b_{k-1}{(2 k + 1) (2 k - 3)}/{8 k}$, while
$\tilde{b}_1=4/5$, $\tilde{b}_2=-17/10$, $\tilde{b}_3=1107/224$, and $\tilde{b}_4=-30\,489/1792$.
We have shown these asymptotic limits along with our calculated densities in Fig. \ref{DensNAsym} for both $Z=Z_c$ and 1. The error in 
the asymptotic forms decreases with increasing $r$ at the expected rates. 

To make our results more immediately useful for
testing density functionals, we created
a fit to the critical density:
\begin{widetext}
\ben
\label{fit}
n_{Z_c}(r)=\left[n_0 e^{-2Z_c r}\left(1+\sum_{k=1}^8\frac{c_k}{2^{k^2/4}}\, r^{k+1}\right)+
B\frac{r^3}{s^3+r^3}\left(\frac{a}{\tilde{y}}\right)^3 \frac{e^{-\tilde{y}}}
{1-\frac{3}{2\tilde{y}}+\frac{21}{8\tilde{y}^2}-
\frac{87}{16\tilde{y}^3}+\frac{1755}{128 \tilde{y}^4}}\right]
\left[1+\frac{d_1r^{10}}{1+d_2r^{29/2}}\right]
\een
\end{widetext}
where $\tilde{y} = (2y)^4/\sqrt{1+(2y)^6}$, $a=4\sqrt{2|\xi_c|}$, $n_0=2\langle \delta(r_1)\rangle$, and 
the fit parameters ($c_k$'s and $d_k$'s) are given in Tab. \ref{FitParameters}.
The short-range part is exact to first order in $r$ and along with the long-range part
contains higher order corrections to order $r^9$
by fitting to the pseudospectral density. The long-range part is chosen to reproduce Eq. (5) to order $r^{-2}$,
while the last term is a Pad\'{e} approximate to the remaining error. For a more accurate density, we provide
our raw data in the supplemental information.
\begin{table}
\begin{tabular}{|c|c|c|c|}
\hline 
$n_0$&$0.23819008067$& $B$&$0.1375$\\
$c_1$&$ 0.0610986$&$c_2 $&$ 0.0352145$\\
$c_3 $&$  -0.0494222$&$c_4 $&$ 0.123575$ \\
$c_5 $&$ -0.212456$&$c_6 $&$ 0.308266$\\
$c_7 $&$ -0.328053$&$c_8 $&$ 0.219550$\\
$d_1 $&$ 7.82582\times 10^{-5}$&$d_2 $&$ 3.79484$\\
$s$&$4.19599$&$^{}$&\\
\hline
\end{tabular}
\caption{\label{FitParameters} Parameters for Eq. (\ref{fit}).}
\end{table}

\begin{table}
\begin{tabular}{|c|c|c|}
\hline
  		&Pseudospectral&fit\\
\hline
$N$ & 1.9999118&1.99757\\
$E$ & -0.414\,986\,212\,52(5)&\\
$E\H$ & 0.595\,467(52)&0.595\,038\\
$T\s$ & 0.389\,857(17)&0.389\,873\\
$E_n$ & -1.053\,346\,537(20)&-1.053\,176\\
$E\c$ & -0.049\,240(39)&-0.049\,202\\
$T\c$ & 0.025\,129(17)&0.0251\,133\\
\hline
\end{tabular}
\caption{\label{EnergyData} Normalization and energy components (total, Hartree, Kohn-Sham kinetic, nuclear, correlation, and kinetic correlation) of the critically bound system.}
\end{table}

\begin{table}
\begin{tabular}{|c|c|c|}
\hline
  		&$Z=Z_c$&$Z=2$ Ref.\,\cite{GrabowskiChernoff2011}\\
\hline
 $\langle r_1^2\rangle$ & 39.779\,95(20) & 1.193\,482\,995\,30(16)\\
 $\langle r_{12}^2\rangle$ & 81.303\,37(40) & 2.516\,439\,313\,8(6)\\
 $\langle \bd{r}_1\cdot \bd{r}_2\rangle$ & -0.871\,728\,2(66) & -0.064\,736\,661\,60(25)\\
$\langle r_1 \rangle$ & 4.146\,972\,44(58)& 0.929\,472\,295\,02(6))\\
 $\langle r_{12}\rangle$ & 7.083\,427\,6(12) & 1.422\,070\,255\,93(38)\\
 $\langle 1/r_1\rangle$ & 0.578\,108\,619(11) & 1.688\,316\,800\,5(6)\\
 $\langle 1/r_{12}\rangle $ & 0.223\,374\,112(19) & 0.945\,818\,448\,5(6)\\
 $\langle 1/r_1^2\rangle $ & 0.873\,035\,760\,4(46) & 6.017\,408\,866\,1(36)\\
 $\langle 1/r_{12}^2 \rangle $ & 0.085\,788\,151\,9(80) & 1.464\,770\,922\,4(15)\\
$ \langle 1/r_1r_2\rangle$ & 0.239\,016\,167(21)& 2.708\,655\,473\,6(20)\\
$\langle 1/r_1r_{12}\rangle$ & 0.154\,038\,646(14) & 1.920\,943\,921\,1(13)\\
$\langle \delta(r_1)\rangle$ & 0.157\,506\,390\,55(31) & 1.810\,429\,318\,2(12)\\
$\langle \delta(r_{12})\rangle$ & 0.001\,473\,985\,59(13) & 0.106\,345\,370\,53(33)\\
\hline
\end{tabular}
\caption{\label{ExpVals} Expectation values in atomic units.}
\end{table}

For two unpolarized electrons, the ground-state energy and all KS energy components
can be extracted directly from the density and external potential without solving
an interacting problem\,\cite{UG94}.  We perform this procedure here as a test of the accuracy
of our densities.
These energies are listed in Table \ref{EnergyData} for $Z=Z_c$ for both our pseudospectral
density and our parametrized form [Eq. (\ref{fit})].
The errors in this form are  $\sim 0.1\%$ or less.  Thus this fit can be used to test approximate
functionals on, if this level of accuracy is sufficient.
In the supplemental info, we give
tables of more accurate densities. 

We also give expectation values of some simple operators
in Tab.\,\ref{ExpVals}, compared to those for the helium atom. We can see that
at the critical value, the two-electron atom is much fatter than for $Z=2$.
Furthermore, we can determine that it is much more likely that the two electrons
are on opposite sides of the nucleus than for $Z=2$ from the expectation value of $\bd{r}_1\cdot\bd{r}_2$,
which has a value more than an order of magnitude greater.

To conclude, we discuss whether this system ought to be considered {\em strongly
correlated}, as in Ref.\,\cite{MUMG14}.  By several naive criteria, we would say that it is.  The fact
that standard density functional approximations fail so badly for the correlation energy
is one.  In fact, if performed self-consistently, such calculations lose a fraction
of an electron to the aether\,\cite{KSB13}.  
Another is the fact that, in a Hartree-Fock calculation, this system would be
unbound because its HF energy is above that of the single ion.  Finally, the ratio of correlation
to exchange energy, and of kinetic correlation to correlation energy are both smallest for $Z=Z_c$.
For weakly correlated systems, that ratio is almost 1, whereas here it is close to 1/2.

\begin{figure}[!tbp]
\includegraphics[width=1\linewidth]{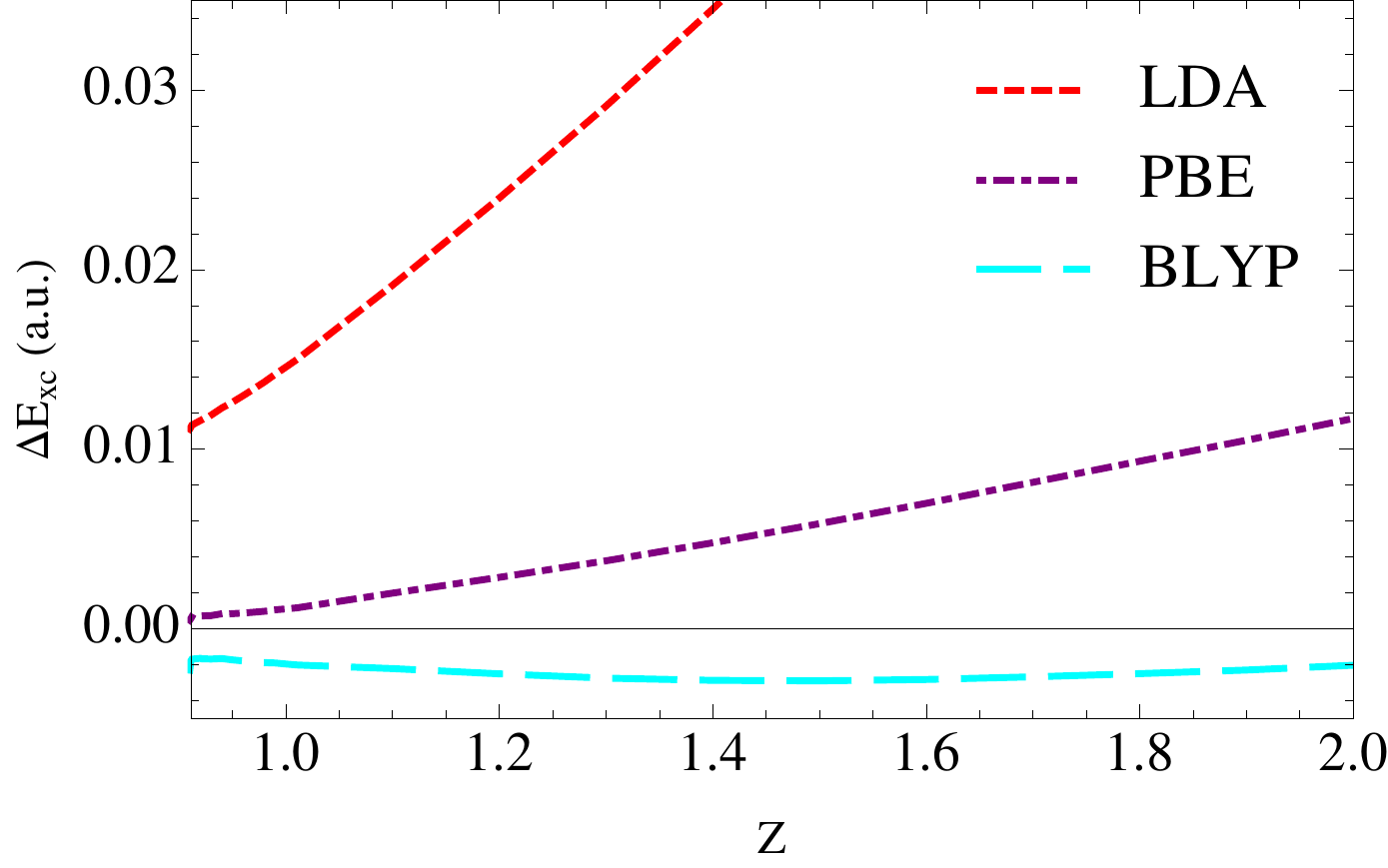}
\caption{\label{dExc} Error in approximate exchange-correlation energies,
evaluated with the exact densities. The error in the PBE energy at $Z=Z_c$ is $6\times 10^{-4}$ hartree.
}
\end{figure}
On the other hand, DFT requires approximating both exchange and correlation together, and is
notorious for cancellations of errors between these two.  
For $Z_c$, not only are the exchange and correlation
potentials qualitatively incorrect in ways similar to Ref.\,\cite{UG94}, 
but also the exchange and correlation energies
are each off by about a factor of two. However, these energy errors almost exactly cancel. The error in the 
exchange-correlation energy using the PBE functional on the exact density is less than a milliHartree (see
Fig. \ref{dExc})! All functionals become more accurate as $Z\to Z\c$, demonstrating the mysterious
power of these approximations.  By this criterion, these are not strongly correlated systems, which explains why
the SCE approach of Ref.\,\cite{MUMG14} does not yield better energetics here.

In summary, we have calculated highly precise wave functions at and near the critical value of $Z$ which has zero ionization
energy for a two-electron atom. We have shown that the correlation energy as a function of $Z$ behaves in a qualitatively 
different manner than standard DFT approximations
near $Z=Z_c$. Asymptotic expressions for the densities at large $r$ to order $r^{-4}$
were derived for all values $Z\ge Z_c$ and fits were given for the coefficient and ionization energies needed for these expressions. 
We gave an accurate fit for the critical density and showed that it reproduces
all DFT energies and many expectation values of our exact numerical expression to about $0.1\%$ or less. This fit
can easily be used by others to check their DFT approximations. Of further use, are the many expectation values we calculate
directly from our wave function. Lastly, we conclude that this critically bound system
is not strongly correlated because its energy is well represented by commonly used GGA's.

We gratefully acknowledge funding from the National Science Foundation (grant number CHE-1112442).

\bibliography{CriticalZ,OscStrength.bib,Master}

\begin{widetext}
\newpage
\end{widetext}
\section{Supplemental Information}

In the main body of the paper, we gave a fit to our density at $Z=Z_c$. This fit gave DFT energies to $0.1\%$ or less. For those 
readers requiring more accuracy, we provide our best calculation of the one-electron density in Tab.\,\ref{Density}. The accuracy
of these data can be inferred from the value of our normalization in Tab.\,II, which is off by $9\times 10^{-5}$. 
The majority of this error comes from the relatively large errors in the tail of the density. Since the weight for the density
at the furthest out nonzero value is about $3\times 10^6$, the error in the density values is about $3\times 10^{-11}$,
comparable to the error in our value of $Z_c$.

In order to obtain a functional form that can be easily evaluated off the grid or used to calculate derivatives, one should
construct the Cardinal functions:
\ben
C_j(x)=\prod_{k=1,k\ne j}^{N}\frac{x-x_k}{x_j-x_k},
\een
where $x=(1-Z_c r)/(1+ Z_c r)$ and $N=52$, which have the property of being equal to one at one grid point and zero at all the others.
The density at an arbitrary value of $x$ can then be obtained by
\ben
n(x)=\sum_{k=1}^N n_jC_j(x),
\een
the explicit values used for $n_j$ are in Tab.\,\ref{Density} (note, values in the table are divided by $Z_c^3$).

Integration can be performed via quadrature. The values of $x_j$ are the Gaussian quadrature points (roots of the 53$^{\textrm{rd}}$
order Legendre polynomial). For example,
\ben
\int d\bd{r} n(r) = \frac{1}{Z_c^3}\sum_{j=1}^N w_j n_j,
\een
where the values of $w_j$ can be found in Tab.\,\ref{Density}.
Note, that the volume element $4\pi r^2 dr$ and the conversion from the $r$-coordinate to $x$-coordinate have already been taken
into account by the values of $w_j$. Since the weights get rather large when $r_j$ is big and the precision of the density is not very good in the tail, one should be careful about including such points in the quadrature. Usually some sort of truncation scheme is needed, 
chosen at a value of $r$ such that contributions from larger $r$ should be negligible. 

\begin{table*}
\begin{tabular}{|c|c|c|c|}
\hline
  $r_j$ (a.u.) & $ x_j$ (a.u.)& $n_j/Z_c^3$ (a.u.)&$w_j$ (a.u.)\\
\hline
0.0005759641912909017	&	0.9989511111039503	&	0.314682351915279	&	4.660741235847267e-9	\\ 
0.00303925808565137	&	0.9944775909292161	&	0.3132732840892592	&	3.0300099503007153e-7	\\ 
0.007489495670215863	&	0.9864461956515499	&	0.3107441016739011	&	2.9066843483625716e-6	\\ 
0.013959796221790716	&	0.9748838842217445	&	0.30710436909115857	&	0.000013877890940302966	\\ 
0.02249735397316229	&	0.9598318269330866	&	0.3023689171597749	&	0.00004612340448161165	\\ 
0.03316494107684804	&	0.9413438536413591	&	0.29655769404226007	&	0.00012288064229107963	\\ 
0.046041970051123156	&	0.9194861289164246	&	0.28969610387322686	&	0.00028224923029420383	\\ 
0.061225781649952314	&	0.8943368905344953	&	0.2818154628129142	&	0.0005832308520293945	\\ 
0.07883324279239258	&	0.8659861628460676	&	0.27295353887455004	&	0.0011139006628788277	\\ 
0.09900270777331016	&	0.8345354323267345	&	0.2631551627930131	&	0.0020025658396714803	\\ 
0.12189640224230178	&	0.8000972834304684	&	0.252472896572073	&	0.0034331010464428006	\\ 
0.147703303023036	&	0.7627949951937449	&	0.24096774168832766	&	0.005666131691016764	\\ 
0.17664260478943944	&	0.7227620997499832	&	0.22870986212496724	&	0.009068430900444735	\\ 
0.2089678872961224	&	0.6801419042271677	&	0.21577928855460538	&	0.014153905827555817	\\ 
0.24497212550105105	&	0.6350869776952459	&	0.20226655892230294	&	0.021641025428937748	\\ 
0.28499372130027406	&	0.5877586049795791	&	0.18827323726105366	&	0.032533717316659924	\\ 
0.32942378214164697	&	0.5383262092858274	&	0.17391223683317283	&	0.04823599283798371	\\ 
0.3787149317776135	&	0.48696674569809606	&	0.15930785598077488	&	0.07071540224691103	\\ 
0.43339201631982543	&	0.4338640677187617	&	0.14459541626218037	&	0.10273774745849322	\\ 
0.4940651706915445	&	0.3792082691160937	&	0.12992037421715608	&	0.1482066729138109	\\ 
0.5614458450037321	&	0.32319500343480784	&	0.11543676326001442	&	0.21265904048506942	\\ 
0.6363665691262939	&	0.2660247836050018	&	0.10130481511247609	&	0.30399399290119455	\\ 
0.7198054734468116	&	0.20790226415636606	&	0.0876876171812794	&	0.43355629325257544	\\ 
0.8129169082310087	&	0.14903550860694917	&	0.07474669189818135	&	0.6177628843571673	\\ 
0.9170699472802668	&	0.08963524464890056	&	0.06263644690028392	&	0.8805726039488526	\\ 
1.033897173471482	&	0.029914109797338766	&	0.05149755279322456	&	1.2572818945152056	\\ 
1.1653569976191294	&	-0.029914109797338766	&	0.04144946858019295	&	1.800435561708559	\\ 
1.3138139675134421	&	-0.08963524464890056	&	0.032582558170176246	&	2.5891630981063303	\\ 
1.4821432470208165	&	-0.14903550860694917	&	0.02495051560492449	&	3.744155528864454	\\ 
1.6738679412287918	&	-0.20790226415636606	&	0.018564108415709536	&	5.452097769788892	\\ 
1.8933416121746534	&	-0.2660247836050018	&	0.013387488441818883	&	8.006263376861513	\\ 
2.145993806251467	&	-0.32319500343480784	&	0.009338394879270824	&	11.875328334071545	\\ 
2.4386647296695396	&	-0.3792082691160937	&	0.006293338088208658	&	17.822611142279463	\\ 
2.7800680689846717	&	-0.4338640677187617	&	0.004097904475823055	&	27.117747215953976	\\ 
3.1814412499350295	&	-0.48696674569809606	&	0.0025830926921683018	&	41.92262357320114	\\ 
3.657475177082267	&	-0.5383262092858274	&	0.0015807827534237687	&	66.01620223603643	\\ 
4.2276696497961455	&	-0.5877586049795791	&	0.0009415599134492133	&	106.20157190068788	\\ 
4.9183526634276635	&	-0.6350869776952459	&	0.0005453042282610736	&	175.1405955662702	\\ 
5.76576296728434	&	-0.6801419042271677	&	0.0003049997869381752	&	297.30896094699824	\\ 
6.820887335531765	&	-0.7227620997499832	&	0.00016242082753832506	&	522.1176753124039	\\ 
8.157294259937565	&	-0.7627949951937449	&	0.00008077588981359748	&	954.4527398959269	\\ 
9.884289312564363	&	-0.8000972834304684	&	0.00003665063488089955	&	1830.417963708067	\\ 
12.169963155779914	&	-0.8345354323267345	&	0.000014740319797992315	&	3719.7467675090556	\\ 
15.28364511271759	&	-0.8659861628460676	&	5.054109250994349e-6	&	8117.066837845993	\\ 
19.678953431941505	&	-0.8943368905344953	&	1.3947108874720773e-6	&	19366.157899552323	\\ 
26.168717467688783	&	-0.9194861289164246	&	2.8316551169626493e-7	&	51822.50322060266	\\ 
36.329306394116436	&	-0.9413438536413591	&	3.6357597278623005e-8	&	161516.57680062292	\\ 
53.55560068801603	&	-0.9598318269330866	&	2.2296122675325307e-9	&	622216.7754332445	\\ 
86.30923308485625	&	-0.9748838842217445	&	3.32223667644662e-11	&	3.2798803957197545e6	\\ 
160.8732228413056	&	-0.9864461956515499	&	0.	&	2.880656250507084e7	\\ 
396.4320475486722	&	-0.9944775909292161	&	0.	&	6.724307713324147e8	\\ 
2091.8996773448594	&	-0.9989511111039503	&	0.	&	2.2330122223506357e11	\\ 
 \hline
\end{tabular}
\caption{\label{Density} Density at the pseudospectral grid points and quadrature weights.}
\end{table*}

\label{page:end}
\end{document}